\documentclass[twocolumn,10pt,pra,showpacs,superscriptaddress]{revtex4}
\usepackage[dvipdfm, colorlinks=true, pdfstartview=FitV, linkcolor=red, citecolor=blue, urlcolor=blue]{hyperref}
\usepackage{amssymb}
\usepackage{amsmath}
\usepackage{graphicx}
\usepackage{amsfonts}

\newtheorem{lemma}{Lemma}

\newtheorem{proposition}{Proposition}

\begin{document}

\title{Indirect control with quantum accessor: coherent control by initial
state preparation}
\author{H. Dong}
\affiliation{Institute of Theoretical Physics, Chinese Academy of Sciences, Beijing,
100080, China}
\author{X.F. Liu}
\affiliation{Department of Mathematics, Beijing University, Beijing 100871, China}
\author{H.C. Fu}
\affiliation{School of Physics,Shenzhen University, Shenzhen 518060, China}
\author{C.P. Sun}
\affiliation{Institute of Theoretical Physics, Chinese Academy of Sciences, Beijing,
100080, China}
\date{\today }

\begin{abstract}
This is the second one in our series of papers on indirect quantum control
assisted by quantum accessor. In this paper we propose and study a new class
of indirect quantum control(IDQC) scheme based on the initial states
preparation of the accessor. In the present scheme, after the initial state
of the accessor is properly prepared, the system is controlled by repeatedly
switching on and off the interaction between the system and the accessor.
This is different from the protocol of our first paper, where we manipulate
the interaction between the controlled system and the accessor. We prove the
controllability of the controlled system for the proposed indirect control
scheme. Furthermore, we give an example with two coupled spins qubits to
illustrate the scheme, the concrete control process and the controllability.
\end{abstract}

\maketitle


\section{Introduction}

Quantum control is a coherent manipulation of a quantum system,
which enables a time evolution from an arbitrary initial state to
arbitrarily given target states
\cite{Tarn,book1,book2,book3,Warren1993,rabitz1996,viola1999,rabitz2000,lloyd2000,roa,tarn2007}.
Recently, we proposed the conception of indirect quantum
control(IDQC) \cite{xue,ind3}. A similar
work under the name of \emph{incoherent} quantum control \cite%
{ind1,ind2,ind4} was proposed by R. Roman et.al. for the control of
spin-half particles. The IDQC is a coherent manipulation for a quantum
system via a coupled intermediate system (called quantum accessor), which
can be controlled directly. Through the engineered interaction between the
controlled quantum system and the accessor, this indirect manipulation
enable an ideal control of the system. A similar controllability for
incoherent control \cite{Tarn} has been studied in the $SU(2)\otimes SU(2)$
case that both the controlled system and quantum accessor are spin-$1/2$
particles \cite{ind1,ind2}.

Our first paper \cite{ind3} on indirect quantum control was motivated by the
pervious works on the quantum control with built-in feedback \cite{xue} and
on the dipole control for finite quantum systems \cite{fu1,fu2}. In Ref.
\cite{ind3}, a general indirect quantum control protocol is proposed and
studied in detail with an engineered interaction totally fixed between the
controlled quantum system and the accessor. In this scheme the accessor is
modeled as a qubit chain with XY-type coupling. Conditions for the complete
controllability, such as the minimal number of qubits in accessor, the
coupling way between the system and the accessor, are investigated in detail.

Different from the scheme proposed in Ref. \cite{ind3}, the quantum control
scheme proposed in this paper is mainly based on ingeniously preparing the
initial state of the quantum \textquotedblleft $\emph{accessor}$%
\textquotedblright . In this scheme, the control function is encoded in a
series of initial states of the accessor, which is prepared by means of a
classical field, and realized through the conditional evolution of the
controlled system from the initial states. So it is adequate to say that
this is essentially an indirect control protocol where the initial
preparation of the \textquotedblleft quantized accessor" determines the time
evolution of the controlled system to realize the target state.

In this paper, we assume that the interaction which governs the evolution of
the controlled system depends on the initial state of the accessor, and it
is switched off when the external field is switched on to manipulate the
accessor. This \textquotedblleft switch off and on
duality\textquotedblright\ can be approximately realized when the strength
of the external field is much stronger than the system-accessor coupling.
Actually, in physical situations it is rather difficult to control the
system state directly, but it is easy to manipulate the accessor state and
modify the $\emph{system}$ state by switching on and off the interaction
between them. We will only consider the quantum controllability of finite
dimensional systems, and leave the relevant infinite case \cite%
{Tarn-infinite} as an open problem.

\begin{figure}[tbp]
\includegraphics[bb=56 233 517 651, width=7 cm, clip]{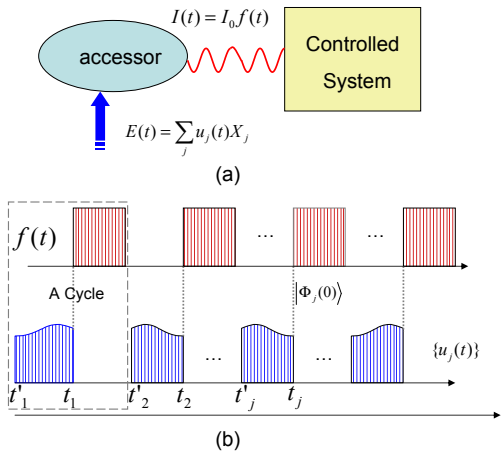}
\caption{(Color online) Illustration of indirect quantum control. (a) The
system is consisted of the controlled system and the accessor. An external
field is utilized to prepare the initial state of the accessor. (b) First
the interaction is closed and the accessor $A$ initialized in the eigenstate
of its free Hamiltonian. Then the interaction is opened to govern the
evolution of $A$ and $S$. The above two steps complete a control cycle. The
cycle is repeated until the target state is obtained.}
\label{Fig:controlscehme}
\end{figure}

The remainder of this paper is organized as follows. In Sec. \ref{csheme},
we describe in detail the control steps of our IDQC scheme by a model. In
Sec. \ref{controllability}, we generally investigate the controllability, in
a slightly generalized sense, of the scheme along the line of Ref. \cite%
{jurdjeiv}. In Sec. \ref{example}, we consider explicitly an example to
illustrate the IDQC scheme and the concerned controllability.

\section{Control scheme}

\label{csheme}

Let the controlled system $S$ be a finite-level system described by the
Hamiltonian $H_{S}$ and the quantum accessor $A$ be a high-dimension system
described by the Hamiltonian $H_{A}$. In our scheme the control process is
steered by the preparation of the initial states of $A$. To realize the
conditional evolution of the system, we require the interaction $H_{I}(S,A)$
between the system and the accessor commute with $H_{A}$, i.e., $\left[
H_{A},H_{I}\left( S,A\right) \right] =0$. Thus the Hamiltonian of the total
system of $S$ and $A$ can be written as
\begin{equation}
H=H_{S}+H_{A}+H_{I}(S,A)\text{.}
\end{equation}%
We observe that in our model $H_{A}$ commutes with both $H_{S}$ and $%
H_{I}(S,A)$. Actually, in the \textquotedblleft standard\textquotedblright\
approach for ideal quantum measurement, we require a non-demolition
property, defined by the above commutation relations. Here we can imagine
the controlled system as a detector, which will not dissipate the energy of
the accessor, but on the other hand can record the manipulation information
of the accessor. Then the evolution of the controlled system is steered by
the quantum states of $A$. The interaction Hamiltonian $H_{I}(S,A)$ between
the two subsystems enjoys the following nice property. For the fixed
variable of the controlled system, which can be viewed as the classical
parameter, the operator $H_{A}$ and $H_{I}\left( S,A\right) $ have the
common eigenstates $|\phi _{j}\rangle $ because of the commutating relation.
The corresponding eigenvalues $H_{j}\left( S\right) $ are parameterized by
the variables of the controlled system and thus can be regarded as an
operator acting on the Hilbert space of the controlled system. Thus for each
eigenstate $|\phi _{j}\rangle $ of the accessor and an arbitrary state $%
|\psi \rangle $ of the system,
\begin{equation}
H_{I}(S,A)|\psi \rangle \otimes |\phi _{j}\rangle =H_{j}(S)|\psi \rangle
\otimes |\phi _{j}\rangle ,
\end{equation}%
where $H_{j}(S)$ is an operator acting on the state space of $S$ and depends
on the initial eigenstate of the accessor. In this sense, we regard $%
H_{j}\left( S\right) $ as an effective Hamiltonian acting on the controlled
system.

Actually for an arbitrary initial state $|\psi _{0}\rangle $ of $S$ and an
eigenstate $|\phi _{j}\rangle $ Hamiltonian $H_{A}$ of the accessor $A$ with
eigenvalue $\alpha _{j}$, we have
\begin{eqnarray}
&&e^{-i(H_{S}+H_{A}+H_{I}(S,A))t}|\psi _{0}\rangle \otimes |\phi _{j}\rangle
\notag \\
&=&e^{-i(H_{S}+H_{I}(S,A))t}e^{-iH_{A}t}|\psi _{0}\rangle \otimes |\phi
_{j}\rangle  \notag \\
&=&e^{-i(H_{S}+H_{j}(S))t}e^{-i\alpha _{j}t}|\psi _{0}\rangle \otimes |\phi
_{j}\rangle .
\end{eqnarray}%
This means that different initial states of the accessor can induce
different time evolutions of the controlled system, which are defined by the
effective Hamiltonian%
\begin{equation}
H_{\text{{\scriptsize eff}}}\left( j\right) =H_{s}+H_{j}\left( S\right)
-\alpha _{j}.
\end{equation}%
Therefore the different effective Hamiltonian $H_{\text{{\scriptsize eff}}%
}\left( j\right) $ can be induced by preparing the different initial states
of the accessor. The above simple observations motivate us to propose a new
IDQC scheme. In our control scheme, we assume that we can manipulate the
accessor $A$ to any common eigenstate state of $H_{A}$ and $H_{I}(S,A)$. The
system is manipulated by the series of effective Hamiltonians $\{H_{\text{%
{\scriptsize eff}}}\left( j\right) \}$ induced by preparing different
initial states of the accessor. Here, we take the model Hamiltonian as%
\begin{eqnarray*}
H_{\text{total}} &=&H+E\left( t\right) \\
&\equiv &H+\sum_{j}u_{j}\left( t\right) Y_{j},
\end{eqnarray*}%
where $E\left( t\right) $ depends on the variables $Y_{j}$ of the accessor
and $u_{j}\left( t\right) $ is the external field used to prepare the
initial states of the accessor. We assume that $u_{j}\left( t\right) $ is so
strong that the system-accessor coupling can be ignored when the external
field is switched on. Therefore, we can formally use the \textquotedblleft
step\textquotedblright\ function $f\left( t\right) $ to describe the
periodic control scheme. Fig. \ref{Fig:controlscehme} illustrates this
\textquotedblleft switch on-off\textquotedblright\ duality control scheme.
Here the basic element is what we call control cycle (Dashed rectangular
area in Fig. \ref{Fig:controlscehme}) consisting of the following two steps:

\begin{enumerate}
\item[\textbf{S1.}] In the time interval $\left( t_{1}^{\prime
},t_{1}\right) $, the interaction is broken between the accessor and the
controlled system, $f\left( t\right) =0$, and the accessor is prepared on
the state $|\phi _{j_{1}}\rangle $ with the classical field $E\left(
t\right) $.

\item[\textbf{S2.}] In the time interval $\left( t_{1},t_{2}^{\prime
}\right) $, the interaction between $S$ and $A$ is switched on $f\left(
t\right) \neq 0 $ and the classical field is removed. In this step, the
system evolves according to the effective Hamiltonian $H_{\text{{\scriptsize %
eff}}}\left( j\right)$. The encoding control information is
\textquotedblleft recorded\textquotedblright\ by the system and we finish
one cycle of manipulation.
\end{enumerate}

In our scheme, the controllability is realized by repeating this cycle.

\section{Complete controllability of indirect control}

\label{controllability}

In this section, we consider the complete controllability of the controlled
system $S$ with the above indirect scheme. For simplicity, we only consider
the case where the controlled system $S$ is an $N$-level system. To explore
the controllability, we define the set of discrete evolution operators
\begin{eqnarray}
\mathcal{G}=\{\hspace{-0.4cm}&&\prod_{j=1}^{M}e^{-i(H_{S}+H_{j}(S))\Delta
t_{j}}  \notag \\
&&|\Delta t_{j}\geq 0,\,\mbox{and}\,\Delta t_{j}\neq 0\,%
\mbox{for
finitely many}j^{\prime }s\},
\end{eqnarray}%
each element of which represents a manipulation completed in $M$ control
cycles. In the $jth $ cycle, the interaction between $S$ and $A$ is switched
on for $\Delta t_{j}$. The set of attainable states from the state $\phi
_{0} $ is
\begin{equation}
\mathcal{G}_{\phi _{0}}=\{g|\phi _{0}\rangle |g\in \mathcal{G}\},
\end{equation}%
where $|\phi _{0}\rangle $ is the initial state of the controlled system. By
definition, $\mathcal{G}$ is a semigroup and a subset of the Lie group $%
SU(N) $. Mathematically, the subsystem $S$ is completely controllable if $%
\mathcal{G}$ is the whole Lie group $SU(N)$. However, physically speaking,
this condition is by far too strong. It would be well acceptable to call $S$
completely controllable if an arbitrary state of $S$ can be approached by
the state in $\mathcal{G}_{\phi _{0}}$ to an arbitrary precision. This
latter condition means that the closure $\mathcal{\bar{G}}$ of $\mathcal{G}$
in $SU(n)$ is just $SU(N)$ itself. In view of this consideration, we will
investigate $\mathcal{\bar{G}}$.

As far as complete controllability is concerned it is meaningful to
investigate under what conditions $\mathcal{\bar{G}}$ is equal to $SU(N)$.
To this end we need to go into the details of $H_{S}$ and $H_{j}(S)$. But we
will content ourself with considering a special case, which illustrates the
main idea. We assume that for each $j$ the Hamiltonian $H_{j}(S)$ takes the
form $H_{j}(S)=\lambda (\alpha _{j})X_{j}$, where $\alpha _{j}$ is a
parameter that can be manipulated freely to enable $\lambda(\alpha _{j}) $
to take zero or some real fixed number. For convenience, the model with this
property will be referred to as the simplified model. And $|\lambda (\alpha
_{j})|$ can be arbitrarily large, if necessarily. This fact turns out to be
crucial for the complete controllability.

Under the above assumption, the effective Hamiltonian reads%
\begin{equation}
H_{\text{{\scriptsize eff}}}=H_{s}+\sum_{j}\lambda \left( \alpha _{j}\right)
X_{j}-\alpha _{j}.
\end{equation}%
This is exactly the Hamiltonian broadly discussed in the direct quantum
control \cite{jurdjeiv,Tarn}. But here it appears naturally in our scheme.
So in some sense it can be said that the scheme proposed in this paper is at
the bottom of the broadly discussed direct quantum control scheme.
Especially, the method of controllability proving developed in direct
quantum control theory applies in our present case.

The following two lemmas are essentially taken from Ref. \cite{jurdjeiv}. We
would rather omit the proof. But we would like to point out that in the
proof of Lemma 2 the condition "$|\lambda (\alpha _{j})|$ can be arbitrarily
large, if necessarily" plays a crucial role.

\begin{lemma}
The closure $\mathcal{\bar{G}}$ of $\mathcal{G}$ is a Lie subgroup of $SU(N)$%
. \label{lemma1}
\end{lemma}

\begin{lemma}
In the simplified model, for each $j$ the single parameter subgroup $\exp
\left( -itX_{j}\right) $ of $SU(n)$ belongs to $\mathcal{\bar{G}}$. \label%
{Lemma2}
\end{lemma}

\begin{proposition}
For the simplified model, if the Lie algebra $su(N)$ of $SU(N)$ can be
generated by $iH_{S}$ and $iX_{j}$'s, then $\mathcal{\bar{G}}=SU(N)$.
\end{proposition}

\textbf{Proof. }Denote by $\mathcal{L}$ the Lie algebra of $\mathcal{\bar{G}}
$, which is a Lie subgroup of $SU(N)$ according to Lemma 1. It directly
follows from Lemma 2 that $iX_{j}\in \mathcal{L}$ for each $j$. On the other
hand for $\Delta t_{j}\geq 0$
\begin{equation}
e^{-i\Delta t_{j}(H_{S}+\lambda (\alpha _{j})X_{j})}\in \mathcal{G}\subseteq
\mathcal{\bar{G}}
\end{equation}%
by definition. But $\mathcal{\bar{G}}$ is a group, this should be true for
an arbitrary $\Delta t_{j}$. Consequently, we have $-i(H_{S}+\lambda (\alpha
_{j})X_{j})\in \mathcal{L}$, and hence $iH_{S}\in \mathcal{L}$. Finally, as $%
iH_{S} $ and $iX_{j}$'s generate the Lie algebra $su(N)$ and $SU(N)$ is a
connected Lie group we can conclude that $\mathcal{\bar{G}}=SU(N)$. The
proof of the proposition is thus completed.

We have proved the controllability of the simplified model. In this model,
the control is attained by repeatedly and properly preparing the initial
states of the accessor and the controlled system works as a detector, which
records and reads the manipulation information encoded in the initial states
of the accessor. This is also the main idea of the IDQC proposed in this
paper.

\section{An Example of IDQC}

\label{example}

In this section, we illustrate the above control scheme and its
controllability with an example: a two-level accessor controls one qubit. In
this simple example, we show that the scheme is equivalent to that of the
direct quantum control.

Let the controlled system be a qubit described by the Hamiltonian $%
H_{S}=\omega _{S}\sigma _{S}^{z}$ and the accessor be a two-level system
with the free Hamiltonian $H_{A}=\omega _{A}\sigma _{A}^{z}$. We couple the
controlled system and the accessor with the interaction $H_{I}=g\sigma
_{S}^{x}\otimes \sigma _{A}^{z}$. Thus the total Hamiltonian of the system
reads
\begin{equation}
H=\omega _{S}\sigma _{S}^{z}\otimes 1+\omega _{A}1\otimes \sigma
_{A}^{z}+g\sigma _{S}^{x}\otimes \sigma _{A}^{z}.
\end{equation}

\begin{figure}[tbp]
\includegraphics[bb=14 224 573 778, width=7 cm, clip]{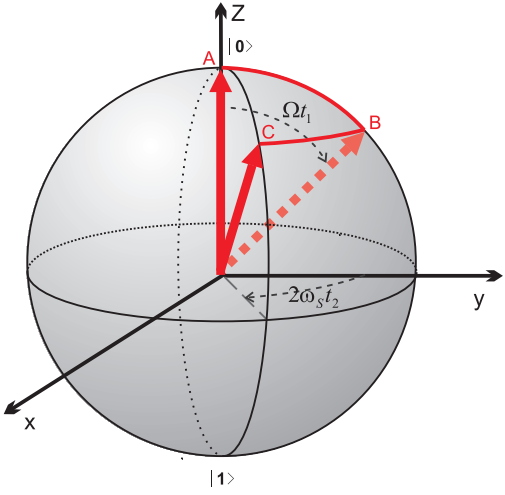}
\caption{(Color online) Illustration of the control protocol in the example.
Firstly, the accessor is prepared on the state $|0\rangle_{A}$ and then the
interaction is switched on for a time interval $t_{1}$. The system evolves
under the effective Hamiltonian. This corresponds to the evolution on the
Bloch sphere along path $\protect\overset{\frown}{AB}$. Secondly, the
interaction is switched off for $t_{2}$ and the system evolves under its
free Hamiltonian. The system evolves along the path $\protect\overset{\frown}%
{BC}$.}
\label{Fig:example}
\end{figure}

It is easily seen that if the initial state of the accessor is $|0\rangle $,
the effective Hamiltonian defined above reads
\begin{equation}
H_{\text{{\scriptsize eff}}}\left( 0\right) =\omega _{S}\sigma
_{S}^{z}-\omega _{A}-g\sigma _{S}^{x}.
\end{equation}%
If the initial state of accessor is $|1\rangle $, we obtain the effective
Hamiltonian%
\begin{equation}
H_{\text{{\scriptsize eff}}}\left( 1\right) =\omega _{S}\sigma
_{S}^{z}+\omega _{A}+g\sigma _{S}^{x}.
\end{equation}%
This effective Hamiltonian is equivalent to the Hamiltonian of a qubit
controlled by a transverse magnetic field the direction of which can be
reversed.

Obviously, we have $i\sigma _{S}^{x}\in su\left( 2\right) $, $i\sigma
_{S}^{z}\in su\left( 2\right) $ and $i\sigma _{S}^{y}=\left[ i\sigma
_{S}^{x},i\sigma _{S}^{z}\right] /2i\in su\left( 2\right) $, producing the
generators of the Lie algebra $SU\left( 2\right) $. Thus when $g>>\omega
_{S} $ Proposition 1 applies and we conclude that the system is controllable.

Next, we would follow the controlling process step by step. The time
evolution operator of the controlled system is calculated straightforward as%
\begin{equation}
U=e^{-i\hat{\varphi}}\left[ \cos \Omega t-i\sin \Omega t\left( \frac{\omega
_{S}}{\Omega }\sigma _{S}^{z}\otimes 1+\frac{g}{\Omega }\sigma
_{S}^{x}\otimes \sigma _{A}^{z}\right) \right] ,
\end{equation}%
where $\Omega =\sqrt{\omega _{S}^{2}+g^{2}}$ and $\hat{\varphi}=\omega
_{A}\sigma _{A}t$. Suppose that the initial state of $S$ is $\left\vert
0\right\rangle _{S}$. First we close the interaction by taking $g=0$ and use
the classical field to prepare the accessor $A$ in the initial state $%
\left\vert 0\right\rangle _{A}$. Then we stop the action of the classical
field and switch on the interaction. Driven by the effective Hamiltonian $H_{%
\text{{\scriptsize eff}}}\left( 0\right) $ for the time interval $t_{1}$,
the quantum state of the controlled system becomes
\begin{equation}
\left\vert \Phi _{0}\right\rangle =e^{i\omega _{A}t}\left[ \alpha \left(
t_{1}\right) \left\vert 0\right\rangle _{S}+\beta \left( t_{1}\right)
\left\vert 1\right\rangle _{S}\right] \otimes \left\vert 0\right\rangle _{A},
\end{equation}%
where $\alpha \left( t_{1}\right) =\left( \cos \Omega t_{1}+i\frac{\omega
_{S}}{\Omega }\sin \Omega t_{1}\right) $ and $\beta \left( t_{1}\right) =i%
\frac{g}{\Omega }\sin \Omega t_{1}$. Noticing that the wave function of the
total system remains factorized, we obtain the state of the controlled
system after the evolution as
\begin{equation}
\left\vert \Phi _{0}\right\rangle _{S}=e^{i\omega _{A}t}\left[ \alpha \left(
t_{1}\right) \left\vert 0\right\rangle _{S}+\beta \left( t_{1}\right)
\left\vert 1\right\rangle _{S}\right] .  \label{fstate}
\end{equation}%
In this equation , when $g>>\omega _{S}$, we obtain $\alpha \left(
t_{1}\right) \approx \cos \Omega t_{1}$ and $\beta \left( t_{1}\right)
\approx i\sin \Omega t_{1}$. Thus the state of the system is approximately%
\begin{equation}
\left\vert \Phi _{0}\right\rangle _{S}=\cos \Omega t_{1}\left\vert
0\right\rangle _{S}+e^{i\pi /2}\sin \Omega t_{1}\left\vert 1\right\rangle
_{S}.
\end{equation}%
After that, we switch off the interaction and the free evolution of the
system for a time interval $t_{2}$. The state of the system then reads%
\begin{equation}
\left\vert \Phi _{0}\right\rangle _{S}=\cos \Omega t_{1}\left\vert
0\right\rangle _{S}+e^{i\pi /2-2i\omega _{S}t_{2}}\sin \Omega
t_{1}\left\vert 1\right\rangle _{S}.
\end{equation}%
The control protocol here is illustrated in the Fig. \ref{Fig:example}.
Since an arbitrary state of the two-level system can be expresses as $%
\left\vert \Phi \right\rangle _{S}=\cos \theta \left\vert 0\right\rangle
_{S}+\sin \theta e^{i\varphi }\left\vert 1\right\rangle _{S}$, we can
conclude that we can prepare the two level system to arbitrary states by
choosing appropriate $t_{1}$ and $t_{2}$. This proves the controllability
directly.

Before concluding our paper, we point out that there is no limit to the
control scheme, and it is different from the observation in Ref. \cite{xue}
that there is a phase uncertainty due to the standard quantum limit. The
Hamiltonian we use here is non-demolition of the accessor, while the one in
Ref. \cite{xue} is of the system. And it preserves the separability of the
wave function. Therefore, there is no control limit to our scheme.

\section{Conclusion}

In this paper, we propose a general scheme to manipulate the quantum states
of a quantum system by preparing the initial state of the accessor.
Different from our previous approach \cite{ind3}, we control the controlled
system by preparing the state of the accessor. Based on the scheme, we
simplify the model and point out its equivalence to broadly investigated
direct quantum control. And we take advantage of the proof of
controllability of direct quantum control and prove the controllability of
arbitrary finite dimensional quantum system. We investigated the concrete
control process of our indirect control scheme with two-level system as an
example. It is found that the two-level system to reach any given target
state by preparing the initial state of accessor (a qubit)only one time and
the time evolution supplies the relative phase of the target state.

Finally, we point out that it is very important to study the concrete
control protocol of our scheme for arbitrary finite dimensional quantum
systems proposed in this paper. Generalization of our approach to infinite
dimensional quantum systems will be also in consideration.

\section*{Acknowledgement}

This work is supported by the NSFC with grant No.190203018, 10474104 and
60433050, 0675085, and NFRPC with No. 2001CB309310 and 2005CB724508.


\begin{thebibliography}{99}
\bibitem{Tarn} G. M. Huang, T. J. Tarn and J. W. Clark, J. Math. Phys.
\textbf{24,} 2608 (1983).

\bibitem{book1} \textit{Information Complexity and Control in Quantum Physics%
}, edited by A. Blaquiere, S. Dinerand and G. Lochak (Springer, New York,
1987).

\bibitem{book2} A. G. Butkovskiy and Yu. I. Samoilenko, \textit{Control of
Quantum-mechanical Processes and Systems} (Kluwer Academic, Dordrecht, 1990).

\bibitem{book3} V. Jurdjevic, \textit{Geometric Control Theory}, Cambridge
University Press, 1997.

\bibitem{Warren1993} W. S. Warren, H. Rabitz, and M. Dahleh, Science \textbf{259}, 1581
(1993).

\bibitem{rabitz1996}V. Ramakrishna and H. Rabitz, Phys. Rev. A \textbf{54}, 1715
(1996).

\bibitem{viola1999} L. Viola, E. Knill, and S. Lloyd, Phys. Rev. Lett. \textbf{82}, 2417 (1999);

\bibitem{lloyd2000} S. Lloyd, Phys. Rev. A \textbf{62}, 022108 (2000).

\bibitem{rabitz2000} H. Rabitz et al., Science \textbf{288}, 824 (2000).

\bibitem{roa} L. Roa, A. Delgado, M. L. Ladronde-Guevara, and A. B. Klimov,
Phys. Rev. A \textbf{73}, 012322 (2006)

\bibitem{tarn2007} N. Ganesan and T. J. Tarn, Phys. Rev. A \textbf{75}, 032323
(2007).

\bibitem{xue} Fei Xue, S.X. Yu, C.P. Sun Phys. Rev. A \textbf{73}, 013403
(2006).

\bibitem{ind3} H. C. Fu, Hui Dong, X. F. Liu and C.P. Sun, Phys. Rev. A
\textbf{75}, 052317 (2007).

\bibitem{ind1} R. Romano and D. D'Alessandro, Phys. Rev. A \textbf{73},
022323 (2006).

\bibitem{ind2} R. Romano and D. D'Alessandro, Phys. Rev. Lett.\textbf{\ 97},
080402 (2006).

\bibitem{ind4} R. Romano, arXiv:0709.1675; R. Romano, arXiv:0707.3383.



\bibitem{fu1} H. Fu, S. G. Schirmer and A. I. Solomon, J. Phys. A. \textbf{34%
}, 1679 (2001).

\bibitem{fu2} S. G. Schirmer, H. Fu and A. I. Solomon, Phys. Rev. A. \textbf{%
63}, 063410 (2001).

\bibitem{Tarn-infinite} R.-B. Wu, T.-J. Tarn, and C.-W. Li, Phys. Rev. A
\textbf{73}, 012719 (2006).

\bibitem{jurdjeiv} V. Jurdjevic and H. J. Sussmann, J. Diff. Eq. \textbf{12}%
, 313 (1972).
\end{thebibliography}
\end{document}